\documentclass[
letterpaper,
linenumbers,
10,
jpb,
]{iopart}

\expandafter\let\csname equation*\endcsname\relax

\expandafter\let\csname endequation*\endcsname\relax

\usepackage{amsmath} 
\usepackage{amssymb} 
\usepackage{graphicx}
\usepackage{hyperref}
\graphicspath{ {./figure/} }
\usepackage{here}
\usepackage{indentfirst}
\usepackage{dcolumn}
\usepackage{bm}
\usepackage{braket}
\usepackage{hyperref}
\usepackage{txfonts}
\usepackage{ulem}
\usepackage{color}
\usepackage{url}
\usepackage{upgreek}

\usepackage{lineno}
\newcommand*\patchAmsMathEnvironmentForLineno[1]{
  \expandafter\let\csname old#1\expandafter\endcsname\csname #1\endcsname
  \expandafter\let\csname oldend#1\expandafter\endcsname\csname end#1\endcsname
  \renewenvironment{#1}
     {\linenomath\csname old#1\endcsname}
     {\csname oldend#1\endcsname\endlinenomath}}
\newcommand*\patchBothAmsMathEnvironmentsForLineno[1]{
  \patchAmsMathEnvironmentForLineno{#1}
  \patchAmsMathEnvironmentForLineno{#1*}}
\AtBeginDocument{
\patchBothAmsMathEnvironmentsForLineno{equation}
\patchBothAmsMathEnvironmentsForLineno{align}
\patchBothAmsMathEnvironmentsForLineno{flalign}
\patchBothAmsMathEnvironmentsForLineno{alignat}
\patchBothAmsMathEnvironmentsForLineno{gather}
\patchBothAmsMathEnvironmentsForLineno{multline}
}
\begin{document}
\title[{}]
{A multiple scattering theoretical approach to time delay in high energy core-level photoemission of heteronuclear diatomic molecules}
\author{
Y Tamura$^{1}$,
K Yamazaki$^{2}$,
K Ueda$^{3}$ and
K Hatada$^{1}$
}

\address{$^1$ Department of Physics, University of Toyama, 3190 Gofuku, 930-8555 Toyama, Japan}
\address{$^2$ Attosecond Science Research Team, Extreme Photonics Research Group, RIKEN Center for Advanced Photonics, RIKEN, 2-1 Hirosawa, Wako, 351-0198 Saitama, Japan.}
\address{$^3$ Department of Chemistry, Graduate School of Science, Tohoku University, 6-3 Aramaki Aza-Aoba, Aoba-ku, 980-8578 Sendai, Japan}

 \eads{\mailto{hatada@sci.u-toyama.ac.jp}}
 \date{\today}

\begin{abstract}
We present analytical expressions of momentum-resolved core-level photoemission time delay in a molecular frame of a heteronuclear diatomic molecule upon photoionization by a linearly polarized soft x-ray attosecond pulse.
For this purpose, we start to derive a general expression of photoemission time delay based on the first order time dependent perturbation theory within the one electron and single channel model in the fixed-in-space system (atoms, molecules and crystals) and apply it to the core-level photoemission within the electric dipole approximation.
By using multiple scattering theory and applying series expansion, plane wave and muffin-tin approximations, the core-level photoemission time delay $t$ is divided into three components, $t_\mathrm{abs}$, $t_\mathrm{path}$ and $t_\mathrm{sc}$, which are atomic photoemission time delay, delays caused by the propagation of photoelectron among the surrounding atoms and the scattering of photoelectron by them, respectively.
We applied a single scattering approximation to $t_\mathrm{path}$ and obtained $t_\mathrm{path}^{(1)}(k,\theta)$ for a linearly polarized soft x-ray field with polarization vector parallel to the molecular axis for a heteronuclear diatomic molecule, where $\theta$ is the angle of measured photoelectron from the molecular axis.
The core-level photoemission time delay $t$ is approximated well with this simplified expression $t_\mathrm{path}^{(1)}(k,\theta)$ in the high energy regime ($k\gtrsim 3.5\,\, \mathrm{a.u.}^{-1}$, where $k$ is the amplitude of photoelectron momentum ${\mathbf k}$), and the validity of this estimated result is confirmed by comparing it with multiple scattering calculations for C 1$s$ core-level photoemission time delay of CO molecules.
$t_\mathrm{path}^{(1)}(k,\theta)$ shows a characteristic dependence on $\theta$, it becomes zero at $\theta=0^{\circ}$, exhibits extended x-ray absorption fine structure (EXAFS) type oscillation with period $2kR$ at $\theta=180^{\circ}$, where $R$ is the bondlength, and gives just the travelling time of photoelectron from the absorbing atom to the neighbouring atom at $\theta=90^{\circ}$.
We confirmed these features also in the numerical results performed by multiple scattering calculations.
\end{abstract}


\submitto{\jpb}
\ioptwocol

Recent developments in attosecond light sources using high-harmonic generation (HHG) combined with the advanced attosecond metrology such as attosecond streaking~\cite{Cavalieri2007,Schultze2010,Pazourek2015} and reconstruction of attosecond beating by interference of two-photon transitions~\cite{Paul2001,Dahlstrom2012} have enabled the measurement of attosecond photoemission time delay in atoms~\cite{Schultze2010,Isinger2017,Peschel2021}, molecules~\cite{Huppert2016,Beaulieu2017,Vos2018,Nandi2020,Biswas2020,Gong2022,Heck2021}, clusters~\cite{Gong2021b}, solids~\cite{Cavalieri2007,Tao2016,Lucchini2016,Ossiander2018}, and liquid~\cite{Jordan2020}, in the range of extreme ultraviolet wavelengths.
The wavelengths of the attosecond HHG light sources have been becoming shorter and shorter and reached the water window region~\cite{Pertot2017,Cousin2017,Saito2019}.
An x-ray free electron laser also has started to provide users with very intense attosecond pulses in the soft x-ray regime~\cite{Duris2019}.
These light source developments open up the door to access the 1$s$ core-level photoemission time delay measurements. 

Here, molecular core-level photoemission time delay in a molecular frame is of particular interest.
Contrary to the photoemission time delay of  spherically symmetric atoms~\cite{Peschel2021}, molecular photoemission time delay exhibits a complex emission angle dependence in the molecular frame~\cite{Vos2018,Gong2022,Heck2021,Holzmeier2021,Rist2021,Chacon2014,Hacket2016,Baykusheva2017}, which reflects the anisotropy of the molecular structure and potential. 
In contrast to valence-level photoemission reflecting the overall anisotropic potential~\cite{Vos2018,Gong2022,Heck2021,Holzmeier2021}, core-level photoemission may give us the atom-resolved potential and geometrical information~\cite{Rist2021} and temporal signature of the entangled nuclear and electronic motion in the photoemission process~\cite{Ghomashi2021}.
Multiple scattering theory~\cite{hatada2010} may be a workhorse to analyze the anisotropic molecular structure and potential as well as nuclear and electronic dynamics hidden in the core-level photoemission time delay in the molecular frame. 

In the present work, we performed a theoretical study on molecular core-level photoemission time delay in the molecular frame.
First, we derived the general expression for the photoemission time delay within the one electron and the single channel model.
Then, introducing electric dipole, muffin-tin~\cite{Natoli2003} and some other approximations described later, we found that core-level photoemission time delay from a unique absorbing atom can be divided into three contributions: (i) atomic photoemission time delay of the absorbing atom $t_\mathrm{abs}$, and photoemission time delays due to (ii) propagation between atoms $t_\mathrm{path}$ and (iii) scattering at each atom $t_\mathrm{sc}$.
We verified that the core-level photoemission time delay asymptotically approaches to $t_{\mathrm{path}}$ within single scattering approximation, $t^{(1)}_{\mathrm{path}}$, in high energy regime ($k\,\,\gtrsim\,\,3.5\,\,\mathrm{a.u.}^{-1}$) by comparing with multiple scattering calculations for C 1$s$ of CO molecules with a linearly polarized x-ray whose polarization vector is parallel to the molecular axis.
We studied the angle dependence of $t^{(1)}_{\mathrm{path}}$.
In particular for perpendicular direction, we found that $t^{(1)}_{\mathrm{path}}$ becomes the travelling time of the photoelectron between the two atoms (i.e. the bond length) in the molecule.

Let us first derive general expressions of the photoemission time delay for any form of the fixed-in-space system upon photoionization by an extreme ultra-violet or x-ray.
Rydberg atomic units are employed in this letter otherwise mentioned.
(Rydberg energy unit $E_{\mathrm{Ryd}}$ is $13.605693$~eV and Rydberg time unit is $\hbar/E_{\mathrm{Ryd}}=48.377687$ atto sec.)

Following the expressions of Goldberger and Watson~\cite{goldberger2004collision}, the photoemission time delay can be defined as the energy derivative of the phase of the photoionization amplitude.
With this in mind, the photoemission time delay in the fixed-in-space system is written with the first order time-dependent perturbation theory within the one electron and the single channel model:
\begin{align}
t\,(\mathbf{k},\hat{\mbox{\boldmath$\varepsilon$}})
\equiv
\frac{1}{2}
\frac{1}{k}
\frac{\mathrm{d}}{\mathrm{d}k}
\,
\mathrm{arg}
\left\{
\bra{\,\psi^{-}(\mathbf{r};\mathbf{k})\,}
\,
e^{\mathrm{i}\boldsymbol{\kappa}\cdot\mathbf{r}}
\hat{\mbox{\boldmath$\varepsilon$}}
\cdot 
\mathbf{r} 
\,
\ket{\,\phi_{ini}(\mathbf{r})}
\right\},
\label{eq:tau}
\end{align}
where $\phi_{ini}$ is the wave function of the initial state, $e^{\mathrm{i}\boldsymbol{\kappa}\cdot\mathbf{r}}\hat{\mbox{\boldmath$\varepsilon$}}\cdot\mathbf{r}$ is the photon field operator in length form with the photon momentum vector $\boldsymbol{\kappa}$ and the polarization vector $\hat{\mbox{\boldmath$\varepsilon$}}=\varepsilon_{1}\,\hat{\mathrm{e}}_{x}+\varepsilon_{-1}\,\hat{\mathrm{e}}_{y}+\varepsilon_{0}\,\hat{\mathrm{e}}_{z}$, and $\psi^{-}(\mathbf{r};\mathbf{k})$ is the wave function of the photoelectron in the final state with momentum vector $\mathbf{k}$ with the incoming boundary condition.
The one half comes from the fact that we use Rydberg atomic units.

We hereafter utilise multiple scattering theory for the photoelectron final state in equation~(\ref{eq:tau})~\cite{hatada2010,Ota2021}.
In the framework of this theory, we partition the space of the target in terms of non-overlapping space-filling cells.
To simplify expressions, we assume that the potential outside the cell domain is zero: namely, interstitial potential $V_0$ is set to 0 and the outersphere is not considered.
Then, the solution of Schr\"{o}dinger equation $\psi^{-}_{i}(\mathbf{r}_{i};\mathbf{k})$ inside site $i$ can be expanded with a set of the angular momentum $L=(l,m)$:
\begin{align}
\psi^{-}_{i}(\mathbf{r}_{i};\mathbf{k})
&=
\sum_{L}
B_{L}^{\,i\,*}(-\mathbf{k})
\,
\bar{\Phi}^{i}_L(\mathbf{r}_{i};k)
\,,
\label{eq:psi-_i}
\\
B^{\,i}_{L}(\mathbf{k})
&=
\sum_{jL'}
\tau_{LL'}^{ij}(k)\,
I_{L'}^{\,j}(\mathbf{k})\,, 
\label{eq:B^i_L}
\\
\tau_{LL'}^{ij} 
&\equiv
\left[\left(T^{-1}-G\right)^{-1}\right]_{LL'}^{ij}
\,,
\label{eq:tauMS}
\\
I_{L}^{\,j}(\mathbf{k})
&\equiv
\mathrm{i}^{\,l}
\sqrt{\frac{k}{\pi}}\, 
e^{\mathrm{i}\mathbf{k}\cdot\mathbf{R}_{jO}}
\mathcal{Y}_{L}(\hat{\mathbf{k}})
\,,
\label{eq:I^i_L}
\end{align}
where $B_{L}^{\,i\,*}(-\mathbf{k})$ is an amplitude and $\bar{\Phi}^{i}_L(\mathbf{r}_{i})$ is the basis function whose more detail is described in Ref.~\cite{hatada2010}.
$i$ and $j$ identify the scattering sites.
$\tau^{ij}_{LL'}$ is a multiple scattering matrix element represented by $T$-matrix and free electron Green's function $G_{LL'}^{ij}$, which is the so-called KKR structure factor~\cite{Natoli2003}.
Note that $T$-matrix is diagonal for site indices, $\delta_{ij}\,T_{LL'}^{i}$, while $G$ is a hallow matrix, $ (1-\delta_{ij}) \,G_{LL'}^{ij}$.
$\mathbf{r}_{i}$ is a position vector from the origin of site $i$ and $\mathbf{R}_{jO}$ is a vector connecting an origin $O$, which is usually a centre of the cluster, to the origin of the scattering site $j$. 
$\mathcal{Y}_{L}(\hat{\mathbf{k}})$ is a real spherical harmonics.
Equation~(\ref{eq:psi-_i}) is interpreted as the following two ways.
One is that the continuum wave function is expanded in the local solution of the local Schr{\"o}dinger equation, $\bar{\Phi}^{i}_L(\mathbf{r}_{i};k)$, which is regular at the origin, and the coefficient $B_{L}^{\,i\,*} (-\mathbf{k})$ is chosen so that $\psi^{-}_{i}(\mathbf{r}_{i};\mathbf{k})$ and its first derivative are smooth over cells.
Another interpretation is that, at a certain time, a photoelectron is excited at the site $i$, then left to the detector from the site $j$ after multiple scattering among sites.
Consequently, the detected photoelectron intensity at a specific angle is the interference of photoelectron waves from all the sites.

Photoionization amplitudes from a core-level of a unique atom $i$ can be described through equation~(\ref{eq:psi-_i}) within electric dipole approximation, $e^{\mathrm{i}\boldsymbol{\kappa}\cdot\mathbf{r}}\sim1$, with setting initial state $\phi_{ini}$ to be the wave function of a core electron $\phi_{L_c}^c$ in the site $i$ with angular momentum quantum number $L_{c}=(0,0)$: 
\begin{align}
\bra{\,\psi^{-}_{i}(\mathbf{r}_{i};\mathbf{k})\,}
\,
\hat{\mbox{\boldmath $\varepsilon$}} 
\cdot 
\mathbf{r}
\,
\ket{\,\phi_{L_{c}}^{\,c}(\mathbf{r}_{i})}
=
\sum_{L}
\sum_{n=-1}^{1}
\varepsilon_{n}\,
M_{L_{c}}^{1nL}(k)\,
B^{\,i}_{L}(-\mathbf{k}) \, ,
\label{eq:amp}
\end{align}
where 
\begin{equation}
M_{L_{c}}^{LL'}(k)
\equiv
\sqrt{\frac{1}{3}}
\int^{R^{i}_{b}}_{0}\mathrm{d}r\,
r^{3}\,
R^{\,*}_{LL'}(r;k)\,
R^{\,c}_{L_{c}}(r)\
\label{eq:M}
\end{equation}
is the transition matrix describing the excitation of the photoelectron and $R_{LL'}$ and $R^{\,c}_{L_{c}}$ are the radial part of the local solution and core wave function, respectively.

Equations~(\ref{eq:B^i_L})-(\ref{eq:I^i_L}) can be further simplified by considering a limit of large $k$ and $R$.
We firstly use series expansion for multiple scattering matrix $\tau$ for a large $kR$ and a spectral radius $\rho(GT)<1$:
\begin{align}
\tau^{ij}_{LL'}
\sim
\delta_{ij}T^{i}_{LL'}
+
\sum_{L_{1}L_{2}}
T^{i}_{LL_{1}}
\left(1-\delta_{ij}\right)
G^{ij}_{L_{1}L_{2}}
T^{j}_{L_{2}L'}
+
\cdots
\,.
\label{eq:SE-tauMT}
\end{align}
Here, the first term is for the direct photoemission from the absorbing atom and the second term corresponds to the single scattering process that the excited photoelectron is propagated to the site $j$ and scattered.
The matrix element of Green's function $G_{LL'}^{ij}$ can be approximated using a plane wave approximation for large $R_{ij}$~\cite{pw1986} as
\begin{align}
G_{LL'}^{ij}
\sim
-4\pi\,
\mathrm{i}^{\,l-l'}\,
\mathcal{G}_{ij}(k)\,
\mathcal{Y}_{L}(\hat{\mathbf{R}}_{ij})\,
\mathcal{Y}_{L'}(\hat{\mathbf{R}}_{ij})\,,
\label{eq:PW-G}
\end{align}
where $\mathcal{G}_{ij}(k)\equiv e^{\mathrm{i}kR_{ij}}/R_{ij}\,$.
For $k$ sufficiently large such as the kinetic energy $\gtrsim 100$ eV, we can employ a muffin-tin approximation for further simplification.
This approximation makes the charge density and potential spherically symmetric at each atom and constant in the interstitial region.
Since this constant potential is chosen to be zero in this application, as mentioned at the beginning, the photoelectron momentum is defined as $k=\sqrt{E}$, where $E$ is the photoelectron kinetic energy.
The radial part of the local solution $R_{LL'}$ and the $T$-matrix $T_{LL'}^{\,i}$ become diagonal by using the muffin-tin approximation.
Then, the amplitude $B^{\,i}_{L}(-\mathbf{k})$ in equation~(\ref{eq:amp}) is simplified using equations~(\ref{eq:SE-tauMT}) and (\ref{eq:PW-G}) as
\begin{align}
B^{\,i}_{L}(-\mathbf{k})
&
\xrightarrow[\substack{\mathrm{muffin-tin}, \\
\mathrm{series\,\,expansion}, \\
\mathrm{plane\,\, wave}}]{}\,
\mathrm{i}^{l}\,
\sqrt{\frac{k}{\pi}}\,\,
T_{l}^{i}(k)\,
e^{-\mathrm{i}{\mathbf{k}}\cdot{\mathbf{R}}_{iO}}
\Biggl\{
\mathcal{Y}_{L}(-\hat{\mathbf{k}})
\nonumber\\
&+
\sum_{j(\neq i)}
\mathcal{Y}_{L}(\hat{\mathbf{R}}_{ij})\,
\mathcal{G}_{ij}(k)\,
f^{j}( k, \theta_{\,\hat{\mathbf{R}}_{ij}, -\hat{\mathbf{k}}} )\,
e^{-\mathrm{i}{\mathbf{k}}\cdot{\mathbf{R}}_{ji}}
 \nonumber\\
&+
\sum_{j_{2}(\neq i)}
\sum_{j_{1}(\neq j_{2})}
\mathcal{Y}_{L}(\hat{\mathbf{R}}_{ij_{1}})\,
\mathcal{G}_{ij_{1}}(k)\,
f^{j_{1}}( k, \theta_{\,\hat{\mathbf{R}}_{ij_{1}}, \hat{\mathbf{R}}_{j_{1}j_{2}}} )
\nonumber\\
&\hspace{1.0cm}
\times
\,\,
\mathcal{G}_{j_{1}j_{2}}(k)\,
f^{j_{2}}( k, \theta_{\,\hat{\mathbf{R}}_{j_{1}j_{2}}, -\hat{\mathbf{k}}})\,
e^{-\mathrm{i}\mathbf{k}\cdot\mathbf{R}_{j_{2}i}}
+
\cdots
\Biggr\}
\nonumber\\
&\hspace{4.5cm}
\equiv \,\,
\overline{B}^{\,i}_{L}
(-\mathbf{k})
\,\,,
\label{eq:MT-B^i_L}
\end{align}
where $f^{\,j}(k,\theta_{\,\hat{\mathbf{r}},\hat{\mathbf{r}}'})\equiv-4\pi\sum_{L}T_{l}^{\,j}(k)\,\mathcal{Y}_L(\hat{\mathbf{r}})\,\mathcal{Y}_L(\hat{\mathbf{r}}')$($\theta_{\,\hat{\mathbf{r}},\hat{\mathbf{r}}'}=\mathrm{arccos}(\hat{\mathbf{r}}\cdot\hat{\mathbf{r}}')$) is the scattering amplitude of site $j$.
In addition, the matrix element in equation~(\ref{eq:M}) is simplified as
\begin{align}
M_{L_{c}}^{L'L}(k)\,
\xrightarrow[\substack{
\mathrm{muffin-tin}\\
\mathrm{approximation}
}]{}
\delta_{L'L}
\sqrt{\frac{1}{3}}
\int^{R^{i}_{b}}_{0}
\mathrm{d}r\,
r^{3}\,
R^{\,*}_{\,l}(r;k)\,
R^{\,c}_{L_{c}}(r)
\nonumber\\
\equiv \,\,
\delta_{L'L}
\overline{M}^{\,l}_{L_{c}}
(k)
\,\,.
\label{eq:MT-M}
\end{align}
Thus, equation~(\ref{eq:amp}) is reduced to the following simple form,
\begin{align}
\bra{\,\psi^{-}_{i}(\mathbf{r}_{i};\mathbf{k})\,}
\,
\hat{\mbox{\boldmath $\varepsilon$}} 
\cdot 
\mathbf{r}
\,
\ket{\,\phi_{L_{c}}^{\,c}(\mathbf{r}_{i})}
\xrightarrow[\substack{
\mathrm{muffin-tin},\\
\mathrm{series\,expansion},\\
\mathrm{plane\,wave}
}]{}
\overline{M}^{\,1}_{L_{c}}(k)
\sum_{n=-1}^{1}
\varepsilon_{n}\,
\overline{B}_{1n}^{\,i}
(-{\mathbf{k}})
\,.
\nonumber\\
\label{eq:MT-amp}
\end{align}
Since the core-level photoemission time delay from a unique absorbing atom at site $i$ is measured with the photoelectron from the origin of the site, it is convenient to set a position of the origin $O$ of the cluster to the one of site $i$ in this application.

Using the approximations described above, the core-level photoemission time delay from a unique absorbing atom at site $i$ can be divided into three components,
\begin{align}
t\,(\mathbf{k},\hat{\mbox{\boldmath$\varepsilon$}})
=
t_\mathrm{abs}(k)
+
t_\mathrm{path}
(\mathbf{k},\hat{\mbox{\boldmath $\varepsilon$}})
+
t_\mathrm{sc}
(\mathbf{k},\hat{\mbox{\boldmath $\varepsilon$}}),
\label{eq:tau3}
\end{align}
where
\begin{align}
t_\mathrm{abs}(k)
&\equiv
\frac{1}{2}
\frac{1}{k}
\frac{\mathrm{d}}{\mathrm{d}k}\,
\mathrm{arg}
\left\{T^{i}_{1}(k)\right\}
\,,
\label{eq:tau_abs}
\\
t_\mathrm{path}(\mathbf{k},\hat{\mbox{\boldmath$\varepsilon$}})
&\equiv
\frac{1}{J^{\,i}(\mathbf{k},\hat{\mbox{\boldmath$\varepsilon$}})}
\frac{1}{2}
\frac{1}{k}\,
\sum_{n,m}\varepsilon_{n}\,\varepsilon_{m}
\nonumber\\
&\times
\mathrm{Im}\,
\Biggl[\,
\overline{B}^{\,i\,*}_{1n}(-{\mathbf{k}})\,
\mathrm{i}
\sqrt{\frac{k}{\pi}}\,\,
T^{i}_{1}(k)\,
\nonumber\\
&\times
\Biggl\{\,
\sum_{j(\neq i)}
\mathcal{Y}_{1m}(\hat{\mathbf{R}}_{ij})\,
f^{j}(k,\theta_{\,\hat{\mathbf{R}}_{ij},-\hat{\mathbf{k}}})\,
\frac{\mathrm{d}}{\mathrm{d}k}
\left(\,
\mathcal{G}_{ij}(k)\,\,
e^{-\mathrm{i}{\mathbf{k}}\cdot{\mathbf{R}}_{ji}}\,
\right)
\nonumber\\
&+
\sum_{j_{2}(\neq i)}
\sum_{j_{1}(\neq j_{2})}
\mathcal{Y}_{1m}(\hat{\mathbf{R}}_{ij_{1}})\,
\frac{\mathrm{d}}{\mathrm{d}k}
\left(\,
\mathcal{G}_{ij_{1}}(k)\,
\mathcal{G}_{j_{1}j_{2}}(k)\,
e^{-\mathrm{i}{\mathbf{k}}\cdot{\mathbf{R}}_{j_{2}i}}\,
\right)
\nonumber\\
\hspace{1.5cm}
&\times
\,
f^{j_{1}}(k, \theta_{\,\hat{\mathbf{R}}_{ij_{1}},\hat{\mathbf{R}}_{j_{1}j_{2}}})\,
f^{j_{2}}(k, \theta_{\,\hat{\mathbf{R}}_{j_{1}j_{2}},-\hat{\mathbf{k}}})
+
\cdots
\, 
\Biggr\}\,
\Biggr]\, ,
\label{eq:tau_path}
\\
t_\mathrm{sc}(\mathbf{k},\hat{\mbox{\boldmath$\varepsilon$}})
&\equiv
\frac{1}{J^{\,i}(\mathbf{k},\hat{\mbox{\boldmath $\varepsilon$}})}
\frac{1}{2}
\frac{1}{k}\,
\sum_{n,m}
\varepsilon_{n}\,\varepsilon_{m}
\nonumber\\
&\times
\mathrm{Im}\,
\Biggl[\,
\overline{B}_{1n}^{\,i\,*}(-{\mathbf{k}})\,
\mathrm{i}\,
\sqrt{\frac{k}{\pi}}\,\,
T^{i}_{1}(k)
\nonumber\\
&\times
\Biggl\{\,
\sum_{j(\neq i)}
\mathcal{Y}_{1m}(\hat{\mathbf{R}}_{ij})\,\,
\frac{\mathrm{d}}{\mathrm{d}k}
\left(\,
f^{j}(k,\theta_{\,\hat{\mathbf{R}}_{ij}, -\hat{\mathbf{k}}})\,
\right)
\mathcal{G}_{ij}(k)\,\,
e^{-\mathrm{i}{\mathbf{k}}\cdot{\mathbf{R}}_{ji}}
\nonumber\\
&+
\sum_{j_{2}(\neq i)}
\sum_{j_{1}(\neq j_{2})}
\mathcal{Y}_{1m}(\hat{\mathbf{R}}_{ij_{1}})\,\,
\mathcal{G}_{ij_{1}}(k)\,\,
\mathcal{G}_{j_{1}j_{2}}(k)\,\,
e^{-\mathrm{i}{\mathbf{k}}\cdot{\mathbf{R}}_{j_{2}i}}
\nonumber\\
\hspace{1.5cm}
&\times
\,
\frac{\mathrm{d}}{\mathrm{d}k}
\left(\,
f^{j_{1}}(k, \theta_{\,\hat{\mathbf{R}}_{ij_{1}}, \hat{\mathbf{R}}_{j_{1}j_{2}}})\,
f^{j_{2}}(k, \theta_{\,\hat{\mathbf{R}}_{j_{1}j_{2}}, -\hat{\mathbf{k}}})\,
\right)
+
\cdots
\, 
\Biggr\} \,
\Biggr],
\
\label{eq:tau_sc}
\end{align}
and
\begin{align}
J^{\,i}(\mathbf{k},\hat{\mbox{\boldmath $\varepsilon$}})
\equiv
\left|
\sum_{n=-1}^{1}
\varepsilon_{n}\,
\overline{B}_{1n}^{\,i}(-{\mathbf{k}})
\right|^{2}\,. 
\label{eq:Ji}
\end{align}
Here, $J^{\,i}(\mathbf{k},\hat{\mbox{\boldmath $\varepsilon$}})$ is an angular-dependent factor proportional to the intensity of the molecular-frame photoelectron angular distributions.
In this formalism, we assumed the optical potential and energy are real for simplicity.
The total photoemission time delay $t\,(\mathbf{k},\hat{\mbox{\boldmath$\varepsilon$}})$ is the time delay of the travelling time from the absorbing atom $i$ to the detector for the photoelectron with a momentum vector $\mathbf{k}$ from the reference of the free electron with the same momentum.
$t_\mathrm{abs}(k)$ is the delay that occurs during the process that a photoelectron escapes from the absorbing atom right after the photoelectron is excited at the absorbing atom.
It does not depend on the photoemission and polarization angles and is half of Wigner time delay~\cite{Pazourek2015,Dahlstrom2012} for a partial wave of $l=1$.
$t_\mathrm{path}(\mathbf{k},\hat{\mbox{\boldmath $\varepsilon$}})$ and $t_\mathrm{sc}(\mathbf{k},\hat{\mbox{\boldmath $\varepsilon$}})$ are time delays that occurred after the photoelectron left the absorbing atom. 
They are collections of $k-$derivatives, the former one is about propagators $\mathcal{G}$ and eigenvalues of translation operator $e^{-\mathrm{i}{\mathbf{k}}\cdot{\mathbf{R}}_{ji}}$, which are related to the geometrical structure of molecules, liquids, or solids, and the latter one is about the scattering amplitudes $f$.
In other words, $t_\mathrm{path}$ is a delay caused by the propagation of photoelectron between atoms in molecules, liquids, or solids, while $t_\mathrm{sc}$ is the one due to scatterings by atoms.

\begin{figure}[htb]
\includegraphics[width=1.0\linewidth]
{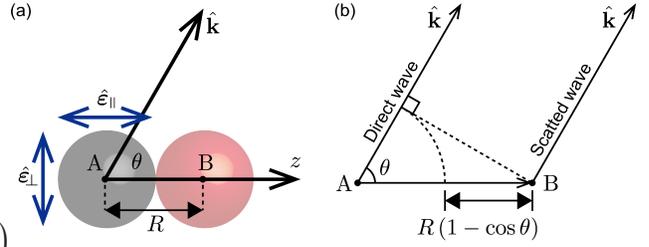}
\caption{
(a)
Definition of the coordinate of a molecule AB in molecular frame:
The molecular axis from A to B is chosen to be parallel to the $z$-axis and $\theta$ is the angle between the momentum vector of photoelectron $\mathbf{k}$ and the molecular axis.
The polarization vector $\hat{\mbox{\boldmath$\varepsilon$}}_{\perp}$ is perpendicular to the molecular axis, while $\hat{\mbox{\boldmath$\varepsilon$}}_{\parallel}$ is parallel.
(b) 
$R(1-\cos \theta)$ is the difference of pathways from A to the detector between the single scattered wave and the reference free wave along the path of the direct wave.
}
\label{fig:coordinate}
\end{figure}

We apply the obtained theoretical model to core-level photoemission from atom A in a heteronuclear diatomic molecule AB in molecular frame, where the molecular axis vector $\mathbf{R}$($\equiv\mathbf{R}_{\mathrm{BA}}$) is parallel to $z$-axis and $\hat{\mathbf{R}} \cdot \hat{\mathbf{k}} = \cos \theta$ (see figure~\ref{fig:coordinate} (a)), by a linearly polarized x-ray.
We consider two cases that the polarization vector is parallel, $\hat{\mbox{\boldmath $\varepsilon$}}_{\parallel}$, and perpendicular, $\hat{\mbox{\boldmath $\varepsilon$}}_{\perp}$, to $\hat{\mathbf{R}}$.
Under the electric dipole approximation, the angular distributions of the direct wave excited with $\hat{\mbox{\boldmath $\varepsilon$}}_{\parallel}$ and $\hat{\mbox{\boldmath $\varepsilon$}}_{\perp}$ are proportional to $\cos\theta$ and $\sin\theta$, respectively.

In the case of $\hat{\mbox{\boldmath $\varepsilon$}}_{\perp}$, we obtain the following simple relations,
\begin{align}
&t\,(\mathbf{k},\hat{\mbox{\boldmath $\varepsilon$}}_{\perp})
=
t_\mathrm{abs}(k)
\, ,
\\
&t_\mathrm{path}
(
 \mathbf{k}, 
 \hat{\mbox{\boldmath $\varepsilon$}}_{\perp}
)
=
t_\mathrm{sc}
(\mathbf{k},\hat{\mbox{\boldmath $\varepsilon$}}_{\perp})
=
0
\, .
\end{align}
These relations reflect the following physical mechanism: since photoelectrons are not emitted in the direction of atom B, no photoelectrons are scattered by atom B.
As a consequence, only photoelectrons directly emitted from atom A are observed.
Therefore, the molecule-specific time delays $t_\mathrm{path}(\mathbf{k},\hat{\mbox{\boldmath $\varepsilon$}}_{\perp})$ and $t_\mathrm{sc}(\mathbf{k},\hat{\mbox{\boldmath $\varepsilon$}}_{\perp})$ become zero, and $t\,(\mathbf{k},\hat{\mbox{\boldmath $\varepsilon$}}_{\perp})$ is reduced to the core-level photoemission delay of a single atom A.

To consider the case of $\hat{\mbox{\boldmath $\varepsilon$}}_{\parallel}$, we employ the single scattering approximation which takes into account up to the second term in equation~(\ref{eq:SE-tauMT}): $\tau\sim T+TGT$.
This works in the high energy limit ($\gtrsim 100$ eV).
According to this theoretical model, the continuum state is the superposition of the direct wave from atom A and the single scattered wave at B after propagation from A to B.
These two waves correspond to the first and second terms in equation~(\ref{eq:MT-B^i_L}), respectively.
Thus, we obtain a compact form for $t_\mathrm{path}^{(1)}(k,\theta)$,
\begin{align}
\label{eq:taupath}
t_\mathrm{path}^{(1)}(k,\theta)
=
\frac{1}{2}
\frac{R\left(1-\cos\theta\right)}{k}
\times
\frac{1}{2}
\left(1-
\frac
{\cos^{2}\theta - \displaystyle{\frac{1}{R^{2}}} \left| f^{\mathrm{B}}(k,\theta) \right|^{2}}
{
J^{\,\mathrm{A}\,(1)}(k,\theta)
}
\right)
\nonumber\\
\end{align}
and
\begin{align}
\label{eq:J}
&
J^{\,\mathrm{A}\,(1)}(k,\theta)
=
\cos^{2}\theta
+
\frac{1}{R^{2}} 
\left| f^{\mathrm{B}} (k,\theta) \right|^{2}
\nonumber\\
&+
2
\cos\theta \,
\frac{1}{R} 
\left| f^{\mathrm{B}}(k,\theta) \right|
\cos \left[ kR \left( 1-\cos \theta \right) + \mathrm{arg}\left\{ f^{\mathrm{B}}(k,\theta) \right\} \right]
~, 
\nonumber\\
\end{align}
where the superscript $(1)$ indicates the scattering order and $J^{\,\mathrm{A}\,(1)}(k,\theta)$ is a first order approximation of equation~(\ref{eq:Ji}) multiplied by $\left|T_{1}^{\mathrm{A}}\right|^{2}k/\pi$.
Hereafter, we rewrite the argument $(\mathbf{k},\hat{\mbox{\boldmath $\varepsilon$}}_{\parallel})$ as $(k,\theta)$.
$t_\mathrm{path}^{(1)}(k,\theta)$ can be interpreted as follows: the factor $R\left(1-\cos\theta\right)/k$ is a travelling time over a distance $R\left(1-\cos\theta\right)$, which is the difference of pathway to the detector between the single scattering wave and free-electron wave (see figure~\ref{fig:coordinate}(b)).
The rest of the terms are responsible for the oscillational structure (oscillating with $kR\left(1-\cos\theta\right)$) due to the interference between the direct and scattered waves.
\begin{figure}[htb]
\includegraphics
[width=\linewidth]
{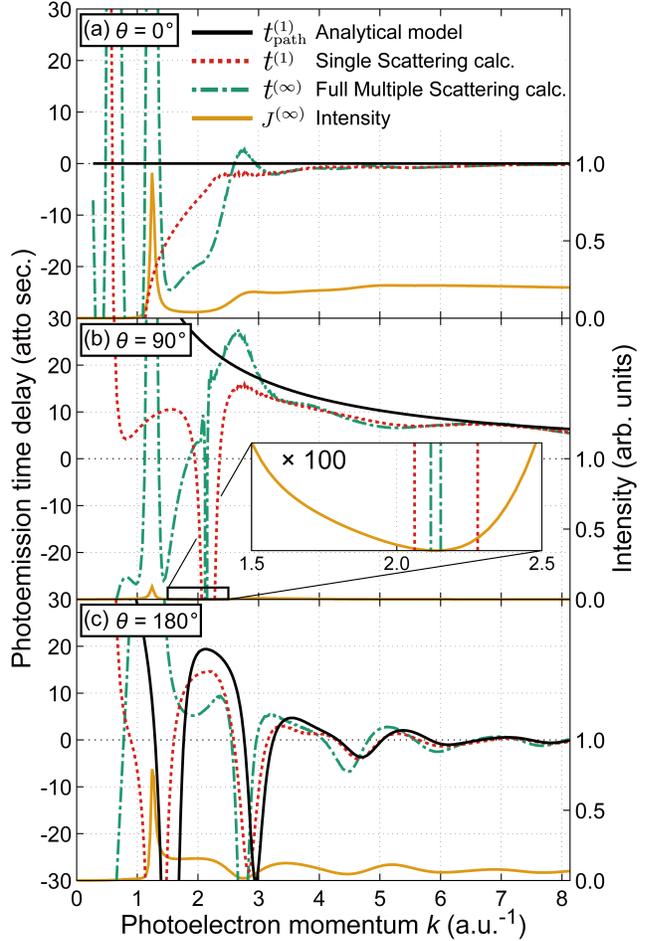}
\caption{
Comparison of the analytical results of $t_\mathrm{path}^{(1)}(k,\theta)$ (black solid line) and the calculation results of total photoemission time delay within single scattering approximation $t^{(1)}(k,\theta)$ (red broken line) and full multiple scattering $t^{(\infty)}(k,\theta)$ (green dash-dotted line) for C 1$s$ core-level photoemission in CO molecule at
(a) $\theta=0^{\circ}$, (b) $90^{\circ}$ and (c) $180^{\circ}$.
The intensity in equation~(\ref{eq:Ji}) calculated within full multiple scattering, $J^{(\infty)}(k,\theta)$, (orange solid line) is shown together in each panel.
The inset is a hundred times magnified zoom.
}
\label{fig:TD}
\end{figure}

We further investigated the photoemission time delay in the molecular frame, using a specific example, i.e. C 1$s$ core-level photoionization in CO.
Since the time delay due to the potential rapidly decreases with the increase in the photoelectron energy, in the high energy regime we expect that $t_{\mathrm{path}}$ ($\sim t^{(1)}_{\mathrm{path}}$) plays a substantial role in equation~(\ref{eq:tau3}).
To verify this, we compared $t^{(1)}_{\mathrm{path}}(k,\theta)$ against more precise forms, i.e. $t^{(\infty)}(k,\theta)$, which is $t(k,\theta)$ considered for the full multiple scattering process, and $t^{(1)}(k,\theta)$ which is its single scattering approximation form (see figure~\ref{fig:TD}).
The calculations for $t^{(1)}(k,\theta)$ and $t^{(\infty)}(k,\theta)$ were performed using the multiple scattering program MsSpec~\cite{sebilleau2011msspec}, taking account of all three terms in equation~(\ref{eq:tau3}).
In figure~\ref{fig:TD}, we see that $t^{(\infty)}(k,\theta)$ converges to $t^{(1)}(k,\theta)$ with increase in energy and both approach $t^{(1)}_\mathrm{path}(k,\theta)$ in the high energy region ($k\,\,\gtrsim\,\, 3.5\,\,\mathrm{a.u.}^{-1}$) at all angles.
Thus, understanding the behaviour of $t^{(1)}_\mathrm{path}(k,\theta)$ helps us to have some insight into the core-level photoemission time delay.

In the low energy region, full multiple scattering (infinite order scattering) should be taken into account.
We see that $t^{(\infty)}(k,\theta)$ diverges to positive values at about $k=1.24$ a.u.$^{-1}$ for each angle in figure~\ref{fig:TD}.
This energy position corresponds to the shape resonance energy, which can be confirmed by the peak position of the intensity in equation~(\ref{eq:Ji}) $J^{(\infty)}(k,\theta)$ calculated with full multiple scattering method.
This large time delay indicates that the photoelectrons are trapped by resonance states originating from the molecular structure.
Although muffin-tin approximation does not work well in the low energy region for highly anisotropic systems such as linear molecules, it is helpful for qualitative understanding of physics.
Another specific phenomenon is that when the intensity becomes zero at $k=0.61$, 2.13 and 2.75 for $\theta=0^{\circ}$, 90$^{\circ}$ and 180$^{\circ}$, respectively, the photoemission time delay becomes large.
It can be understood that this divergence is due to the intensity in the denominator getting to zero in equations~(\ref{eq:tau_path}) and (\ref{eq:tau_sc}).

Let us look into the details of angular dependence of $t^{(1)}_{\mathrm{path}}(k,\theta)$, especially for some specific angles $0^{\circ}$, $90^{\circ}$ and $180^{\circ}$.
At $\theta=0^{\circ}$,
\begin{equation}
t^{(1)}_\mathrm{path}(k,0)=0.
\label{eq:tau0}
\end{equation}
As we see, there is no delay caused.
This is because the single scattering wave follows the same photoemission pathway as the free electron wave.
Thus this equation justifies that the core-level photoemission time delay measured at $\theta=0^{\circ}$ in the high-energy region can be treated as a reference for the measurement.
At $\theta=90^{\circ}$,
\begin{align}
t_\mathrm{path}^{(1)}\left(k, \frac{\pi}{2}\right)
=
\frac{1}{2}
\frac{R}{k}\,.
\label{eq:tau90}
\end{align}
We notice that this is just the travelling time of the photoelectron from atom A to B, so that this relation may be used as a probe of the bondlength of a heteronuclear diatomic molecule.
This can be explained by the following simple considerations.
Since $\theta=90^{\circ}$ is the forbidden angle for direct wave (remember now $\hat{\mbox{\boldmath $\varepsilon$}}_{\parallel}$) under the electric dipole approximation, the detected photoelectron must be scattered by the surrounded atom after propagation from A to B.
Thus, the time delay is caused by the propagation.
At $\theta=180^{\circ}$, $t_\mathrm{path}^{(1)}$ oscillates with frequency $2R$ according to equation~(\ref{eq:J}).
The oscillatory structure is practically the same in origin as the extended x-ray absorption fine structure (EXAFS) and is caused by the interference between the direct photoelectron wave and the scattered wave.
Thus it also includes the structural information.

In conclusion, we first derived expressions of core-level photoemission time delay using multiple scattering theory in equations~(\ref{eq:tau_abs})-(\ref{eq:Ji}).
These equations work for any kind of system, not only molecules but also 2D and 3D extended systems.
We applied our theoretical model to heteronuclear diatomic molecules in the high-energy region.
The results revealed the angular dependence of the photoemission time delay in the molecular frame and provided a physical interpretation of the photoemission time delay caused by the molecular structure, which is not present in the photoemission time delay of a single atom.
In particular, the photoemission time delay observed in the direction perpendicular to the molecular axis set to parallel to the polarization axis can be interpreted as the travelling time spent by photoelectron to propagate between two atoms.
This feature indicates that the measurement of core-level photoemission time delay can be a probe not only for dynamics of molecular photoionization (i.e. photoelectron scattering by the anisotropic molecular potential) but also for structural dynamics, e.g. for structural changes of molecules photo-excited by the ultraviolet laser.
Further extension of this work will provide a new probe for the electronic and structural dynamics of polyatomic molecules, clusters and materials in the solid and liquid phases.

\section*{Acknowledgements}
We would like to thank Dr. F. Ota for fruitful discussion.
K. H. acknowledges funding by JST CREST Grant No. JPMJCR1861. K. Y. is grateful for JSPS KAKENHI Grant Number 19H05628.

\section*{References}
\bibliographystyle{unsrt}
\bibliography{yoshiaki}

\begin{thebibliography}{10}

\bibitem{Cavalieri2007}
A.~L. Cavalieri, N.~M\"{u}ller, Th. Uphues, V.~S. Yakovlev, A.~Baltu{\v{s}}ka,
  B.~Horvath, B.~Schmidt, L.~Bl\"{u}mel, R.~Holzwarth, S.~Hendel, M.~Drescher,
  U.~Kleineberg, P.~M. Echenique, R.~Kienberger, F.~Krausz, and U.~Heinzmann.
\newblock Attosecond spectroscopy in condensed matter.
\newblock {\em Nature}, 449(7165):1029--1032, October 2007.

\bibitem{Schultze2010}
M.~Schultze, M.~Fiess, N.~Karpowicz, J.~Gagnon, M.~Korbman, M.~Hofstetter,
  S.~Neppl, A.~L. Cavalieri, Y.~Komninos, T.~Mercouris, C.~A. Nicolaides,
  R.~Pazourek, S.~Nagele, J.~Feist, J.~Burgdorfer, A.~M. Azzeer, R.~Ernstorfer,
  R.~Kienberger, U.~Kleineberg, E.~Goulielmakis, F.~Krausz, and V.~S. Yakovlev.
\newblock Delay in photoemission.
\newblock {\em Science}, 328(5986):1658--1662, June 2010.

\bibitem{Pazourek2015}
Renate Pazourek, Stefan Nagele, and Joachim Burgd\"{o}rfer.
\newblock Attosecond chronoscopy of photoemission.
\newblock {\em Reviews of Modern Physics}, 87(3):765--802, August 2015.

\bibitem{Paul2001}
P.~M. Paul, E.~S. Toma, P.~Breger, G.~Mullot, F.~Aug\'e, Ph. Balcou, H.~G.
  Muller, and P.~Agostini.
\newblock Observation of a train of attosecond pulses from high harmonic
  generation.
\newblock {\em Science}, 292(5522):1689--1692, June 2001.

\bibitem{Dahlstrom2012}
J~M Dahlstr\"{o}m, A~L'Huillier, and A~Maquet.
\newblock Introduction to attosecond delays in photoionization.
\newblock {\em Journal of Physics B: Atomic, Molecular and Optical Physics},
  45(18):183001, August 2012.

\bibitem{Isinger2017}
M.~Isinger, R.~J. Squibb, D.~Busto, S.~Zhong, A.~Harth, D.~Kroon, S.~Nandi,
  C.~L. Arnold, M.~Miranda, J.~M. Dahlstr\"{o}m, E.~Lindroth, R.~Feifel,
  M.~Gisselbrecht, and A.~L'Huillier.
\newblock Photoionization in the time and frequency domain.
\newblock {\em Science}, 358(6365):893--896, November 2017.

\bibitem{Peschel2021}
Jasper Peschel, David Busto, Marius Plach, Mattias Bertolino, Maria Hoflund,
  Sylvain Maclot, Jimmy Vinbladh, Hampus Wikmark, Felipe Zapata, Eva Lindroth,
  Mathieu Gisselbrecht, Jan~Marcus Dahlstr\"{o}m, Anne L'Huillier, and Per
  Eng-Johnsson.
\newblock Complete characterization of multi-channel single photon ionization,
  2021.

\bibitem{Huppert2016}
Martin Huppert, Inga Jordan, Denitsa Baykusheva, Aaron von Conta, and
  Hans~Jakob W\"{o}rner.
\newblock Attosecond delays in molecular photoionization.
\newblock {\em Physical Review Letters}, 117(9), August 2016.

\bibitem{Beaulieu2017}
S.~Beaulieu, A.~Comby, A.~Clergerie, J.~Caillat, D.~Descamps, N.~Dudovich,
  B.~Fabre, R.~G{\'{e}}neaux, F.~L{\'{e}}gar{\'{e}}, S.~Petit, B.~Pons,
  G.~Porat, T.~Ruchon, R.~Ta\"{i}eb, V.~Blanchet, and Y.~Mairesse.
\newblock Attosecond-resolved photoionization of chiral molecules.
\newblock {\em Science}, 358(6368):1288--1294, December 2017.

\bibitem{Vos2018}
J.~Vos, L.~Cattaneo, S.~Patchkovskii, T.~Zimmermann, C.~Cirelli, M.~Lucchini,
  A.~Kheifets, A.~S. Landsman, and U.~Keller.
\newblock Orientation-dependent stereo {W}igner time delay and electron
  localization in a small molecule.
\newblock {\em Science}, 360(6395):1326--1330, June 2018.

\bibitem{Nandi2020}
S.~Nandi, E.~Pl{\'{e}}siat, S.~Zhong, A.~Palacios, D.~Busto, M.~Isinger,
  L.~Neori{\v{c}}i{\'{c}}, C.~L. Arnold, R.~J. Squibb, R.~Feifel, P.~Decleva,
  A.~L'Huillier, F.~Mart{\'{\i}}n, and M.~Gisselbrecht.
\newblock Attosecond timing of electron emission from a molecular shape
  resonance.
\newblock {\em Science Advances}, 6(31):eaba7762, July 2020.

\bibitem{Biswas2020}
Shubhadeep Biswas, Benjamin F\"{o}rg, Lisa Ortmann, Johannes Sch\"{o}tz,
  Wolfgang Schweinberger, Tom{\'{a}}{\v{s}} Zimmermann, Liangwen Pi, Denitsa
  Baykusheva, Hafiz~A. Masood, Ioannis Liontos, Amgad~M. Kamal, Nora~G. Kling,
  Abdullah~F. Alharbi, Meshaal Alharbi, Abdallah~M. Azzeer, Gregor Hartmann,
  Hans~J. W\"{o}rner, Alexandra~S. Landsman, and Matthias~F. Kling.
\newblock Probing molecular environment through photoemission delays.
\newblock {\em Nature Physics}, 16(7):778--783, May 2020.

\bibitem{Gong2022}
Xiaochun Gong, Wenyu Jiang, Jihong Tong, Junjie Qiang, Peifen Lu, Hongcheng Ni,
  Robert Lucchese, Kiyoshi Ueda, and Jian Wu.
\newblock Asymmetric attosecond photoionization in molecular shape resonance.
\newblock {\em Physical Review X}, 12(1), January 2022.

\bibitem{Heck2021}
Saijoscha Heck, Denitsa Baykusheva, Meng Han, Jia-Bao Ji, Conaill Perry,
  Xiaochun Gong, and Hans~Jakob W\"{o}rner.
\newblock Attosecond interferometry of shape resonances in the recoil frame of
  {CF} 4.
\newblock {\em Science Advances}, 7(49), December 2021.

\bibitem{Gong2021b}
Xiaochun Gong, Saijoscha Heck, Denis Jelovina, Conaill Perry, Kristina
  Zinchenko, and Hans~Jakob W\"{o}rner.
\newblock Attosecond spectroscopy of size-resolved water clusters, 2021.

\bibitem{Tao2016}
Zhensheng Tao, Cong Chen, Tibor Szilv{\'{a}}si, Mark Keller, Manos Mavrikakis,
  Henry Kapteyn, and Margaret Murnane.
\newblock Direct time-domain observation of attosecond final-state lifetimes in
  photoemission from solids.
\newblock {\em Science}, 353(6294):62--67, July 2016.

\bibitem{Lucchini2016}
M.~Lucchini, S.~A. Sato, A.~Ludwig, J.~Herrmann, M.~Volkov, L.~Kasmi,
  Y.~Shinohara, K.~Yabana, L.~Gallmann, and U.~Keller.
\newblock Attosecond dynamical franz-keldysh effect in polycrystalline diamond.
\newblock {\em Science}, 353(6302):916--919, August 2016.

\bibitem{Ossiander2018}
M.~Ossiander, J.~Riemensberger, S.~Neppl, M.~Mittermair, M.~Sch\"{a}ffer,
  A.~Duensing, M.~S. Wagner, R.~Heider, M.~Wurzer, M.~Gerl, M.~Schnitzenbaumer,
  J.~V. Barth, F.~Libisch, C.~Lemell, J.~Burgd\"{o}rfer, P.~Feulner, and
  R.~Kienberger.
\newblock Absolute timing of the photoelectric effect.
\newblock {\em Nature}, 561(7723):374--377, September 2018.

\bibitem{Jordan2020}
Inga Jordan, Martin Huppert, Dominik Rattenbacher, Michael Peper, Denis
  Jelovina, Conaill Perry, Aaron von Conta, Axel Schild, and Hans~Jakob
  W\"{o}rner.
\newblock Attosecond spectroscopy of liquid water.
\newblock {\em Science}, 369(6506):974--979, August 2020.

\bibitem{Pertot2017}
Yoann Pertot, C{\'{e}}dric Schmidt, Mary Matthews, Adrien Chauvet, Martin
  Huppert, Vit Svoboda, Aaron von Conta, Andres Tehlar, Denitsa Baykusheva,
  Jean-Pierre Wolf, and Hans~Jakob W\"{o}rner.
\newblock Time-resolved x-ray absorption spectroscopy with a water window
  high-harmonic source.
\newblock {\em Science}, 355(6322):264--267, January 2017.

\bibitem{Cousin2017}
Seth~L. Cousin, Nicola~Di Palo, B{\'{a}}rbara Buades, Stephan~M. Teichmann,
  M.~Reduzzi, M.~Devetta, A.~Kheifets, G.~Sansone, and Jens Biegert.
\newblock Attosecond streaking in the water window: A new regime of attosecond
  pulse characterization.
\newblock {\em Physical Review X}, 7(4), November 2017.

\bibitem{Saito2019}
Nariyuki Saito, Hiroki Sannohe, Nobuhisa Ishii, Teruto Kanai, Nobuhiro Kosugi,
  Yi~Wu, Andrew Chew, Seunghwoi Han, Zenghu Chang, and Jiro Itatani.
\newblock Real-time observation of electronic, vibrational, and rotational
  dynamics in nitric oxide with attosecond soft x-ray pulses at
  400{\hspace{0.167em}}{\hspace{0.167em}}{eV}.
\newblock {\em Optica}, 6(12):1542, December 2019.

\bibitem{Duris2019}
Joseph Duris, Siqi Li, Taran Driver, Elio~G. Champenois, James~P. MacArthur,
  Alberto~A. Lutman, Zhen Zhang, Philipp Rosenberger, Jeff~W. Aldrich, Ryan
  Coffee, Giacomo Coslovich, Franz-Josef Decker, James~M. Glownia, Gregor
  Hartmann, Wolfram Helml, Andrei Kamalov, Jonas Knurr, Jacek Krzywinski,
  Ming-Fu Lin, Jon~P. Marangos, Megan Nantel, Adi Natan, Jordan~T. O'Neal,
  Niranjan Shivaram, Peter Walter, Anna~Li Wang, James~J. Welch, Thomas J.~A.
  Wolf, Joseph~Z. Xu, Matthias~F. Kling, Philip~H. Bucksbaum, Alexander
  Zholents, Zhirong Huang, James~P. Cryan, and Agostino Marinelli.
\newblock Tunable isolated attosecond x-ray pulses with gigawatt peak power
  from a free-electron laser.
\newblock {\em Nature Photonics}, 14(1):30--36, December 2019.

\bibitem{Holzmeier2021}
F.~Holzmeier, J.~Joseph, J.~C. Houver, M.~Lebech, D.~Dowek, and R.~R. Lucchese.
\newblock Influence of shape resonances on the angular dependence of molecular
  photoionization delays.
\newblock {\em Nature Communications}, 12(1), December 2021.

\bibitem{Rist2021}
Jonas Rist, Kim Klyssek, Nikolay~M. Novikovskiy, Max Kircher, Isabel
  Vela-P{\'{e}}rez, Daniel Trabert, Sven Grundmann, Dimitrios Tsitsonis,
  Juliane Siebert, Angelina Geyer, Niklas Melzer, Christian Schwarz, Nils
  Anders, Leon Kaiser, Kilian Fehre, Alexander Hartung, Sebastian Eckart,
  Lothar Ph.~H. Schmidt, Markus~S. Sch\"{o}ffler, Vernon~T. Davis, Joshua~B.
  Williams, Florian Trinter, Reinhard D\"{o}rner, Philipp~V. Demekhin, and Till
  Jahnke.
\newblock Measuring the photoelectron emission delay in the molecular frame.
\newblock {\em Nature Communications}, 12(1), November 2021.

\bibitem{Chacon2014}
Alexis Chacon, Manfred Lein, and Camilo Ruiz.
\newblock Asymmetry of {W}igner's time delay in a small molecule.
\newblock {\em Physical Review A}, 89(5), May 2014.

\bibitem{Hacket2016}
P~Hockett, E~Frumker, D~M Villeneuve, and P~B Corkum.
\newblock Time delay in molecular photoionization.
\newblock {\em Journal of Physics B: Atomic, Molecular and Optical Physics},
  49(9):095602, April 2016.

\bibitem{Baykusheva2017}
Denitsa Baykusheva and Hans~Jakob W\"{o}rner.
\newblock Theory of attosecond delays in molecular photoionization.
\newblock {\em The Journal of Chemical Physics}, 146(12):124306, March 2017.

\bibitem{Ghomashi2021}
Bejan Ghomashi, Nicolas Douguet, and Luca Argenti.
\newblock Attosecond intramolecular scattering and vibronic delays.
\newblock {\em Phys. Rev. Lett.}, 127:203201, Nov 2021.

\bibitem{hatada2010}
Keisuke Hatada, Kuniko Hayakawa, Maurizio Benfatto, and Calogero~R Natoli.
\newblock Full-potential multiple scattering theory with space-filling cells
  for bound and continuum states.
\newblock {\em Journal of Physics: Condensed Matter}, 22(18):185501, April
  2010.

\bibitem{Natoli2003}
C.~R. Natoli, M.~Benfatto, S.~Della~Longa, and K.~Hatada.
\newblock {X-ray absorption spectroscopy: state-of-the-art analysis}.
\newblock {\em Journal of Synchrotron Radiation}, 10(1):26--42, Jan 2003.

\bibitem{goldberger2004collision}
Marvin~L Goldberger and Kenneth~M Watson.
\newblock {\em Collision theory}.
\newblock Courier Corporation, 2004.

\bibitem{Ota2021}
F~Ota, K~Yamazaki, D~S{\'{e}}billeau, K~Ueda, and K~Hatada.
\newblock Theory of polarization-averaged core-level molecular-frame
  photoelectron angular distributions: I. a full-potential method and its
  application to dissociating carbon monoxide dication.
\newblock {\em Journal of Physics B: Atomic, Molecular and Optical Physics},
  54(2):024003, January 2021.

\bibitem{pw1986}
J.~J. Rehr, R.~C. Albers, C.~R. Natoli, and E.~A. Stern.
\newblock New high-energy approximation for x-ray-absorption near-edge
  structure.
\newblock {\em Physical Review B}, 34(6):4350--4353, September 1986.

\bibitem{sebilleau2011msspec}
Didier S{\'e}billeau, Calogero Natoli, George~M Gavaza, Haifeng Zhao, Fabiana
  Da~Pieve, and Keisuke Hatada.
\newblock Msspec-1.0: A multiple scattering package for electron spectroscopies
  in material science.
\newblock {\em Computer Physics Communications}, 182(12):2567--2579, 2011.

\end{thebibliography}

\end{document}